\documentclass[acus]{JAC2003}

\usepackage{graphicx}
\usepackage{subcaption} %  for subfigures environments
\usepackage{booktabs}
\usepackage{pdflscape}
\begin{document}
\title{LLRF System for the Fermilab PIP-II Superconducting LINAC
\thanks{The authors of this work grant the arXiv.org and LLRF Workshop's International Organizing Committee a non-exclusive and irrevocable license to distribute the article, and certify that they have the right to grant this license.}\thanks{ Work supported by Fermi Research Alliance LLC. Under DE-AC02-07CH11359 with U.S. DOE.}}

\author{P. Varghese\thanks{ varghese@fnal.gov}, B. Chase, E. Cullerton, S. Raman, S. Ahmed, P. Hanlet, D. Klepec \\
Fermi National Accelerator Laboratory (FNAL), Batavia, IL 60510, USA \\
L. Doolittle, S. Murthy, C. Du, K. Penney \\
Lawrence Berkeley National  Laboratory (LBNL), Berkeley, CA 94720, USA\\
C. Hovater, J. Latshaw \\
Jefferson National  Laboratory (JLab),Newport News, VA 23606, USA}

\maketitle

\begin{abstract}
   PIP-II is an 800 MEV superconducting linac that is in the initial acceleration chain for the Fermilab accelerator complex. The RF system consists of a warm front-end with
an ion source, RFQ and buncher cavities along with 25 superconducting cryo-modules  comprised of five different acceleration \(\beta\). The LLRF system for the LINAC has
to provide field and resonance control for a total of 125 RF cavities. 
 The LLRF system design is in the final design review phase and will enter
the production phase next year. The PIP-II project is an international collaboration with various
partner labs contributing subsystems. The LLRF system design for the PIP-II Linac is presented and
the specification requirements and system performance in various stages of testing are described in this paper.
\end{abstract}

\section{INTRODUCTION}
\par
The PIP-II project at Fermilab is a new superconducting linac feeding the existing Booster, Recycler and Main Injector accelerator rings enabling them to provide a 1.2 MW proton beam
over the energy range of 60-120 GeV to drive neutrino research at the Deep Underground Neutrino Experiment (DUNE) and for the Mu2e project[1].  The RF system consists of a warm front-end with
an ion source, RFQ, 4  buncher cavities along with 25 superconducting cryo-modules as shown in Fig.2. The superconducting cryo-modules are comprised of five different types - 1 Half Wave Resonator(HWR, \(\beta\)=0.11), 2 Single Spoke Resonator(SSR1, \(\beta\)=0.22), 7 SSR2(\(\beta\)=0.47), 9 Low Beta (LB650, \(\beta\)=0.61) and 4 High Beta(HB650, \(\beta\)=0.92)[1]. The LLRF system for the LINAC has
to provide field and resonance control for a total of 125 RF cavities. Resonance control for the RFQ is by temperature control of the cooling water whereas the HWR cavities use a pneumatic system based on Helium pressure[10]. The remaining cryo-modules use a stepper motor / piezo tuner combination for slow and fast tuning.
\begin{figure}[b]
\centering
\includegraphics[width=3.0in]{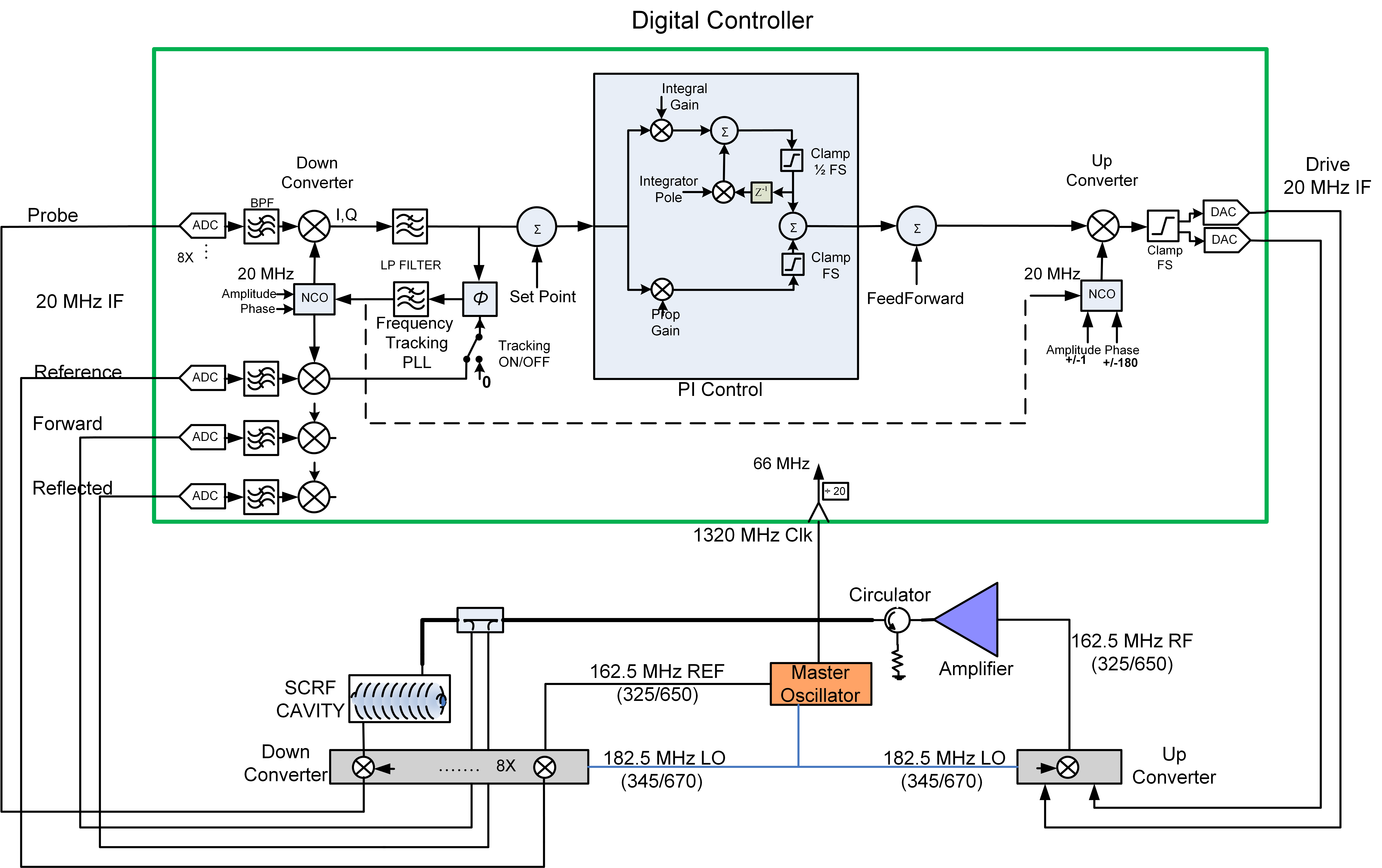}
\caption{LLRF System Architecture -I/Q Control}
\label {fig1}
\end{figure}

\par
The Cryo Module Test Facility(CMTF) is an SRF test facility for the PIP-II project where the injector, warm front-end and the first two superconducting cryomodules were tested[2]. Single cavity testing of all the superconducting cavities
except HWR is an ongoing effort at the Spoke resonator Test Cave(STC) over the past few years. These two test stands have been the primary facilities for testing and characterizing the performance
of the various components of the LLRF systems, RF cavities and other subsystems of the PIP-II linac. 
 The LLRF system design for the linac is described and
the specification requirements and system performance are discussed.
\begin{figure*}[!t]
\centering
\includegraphics[width=165mm]{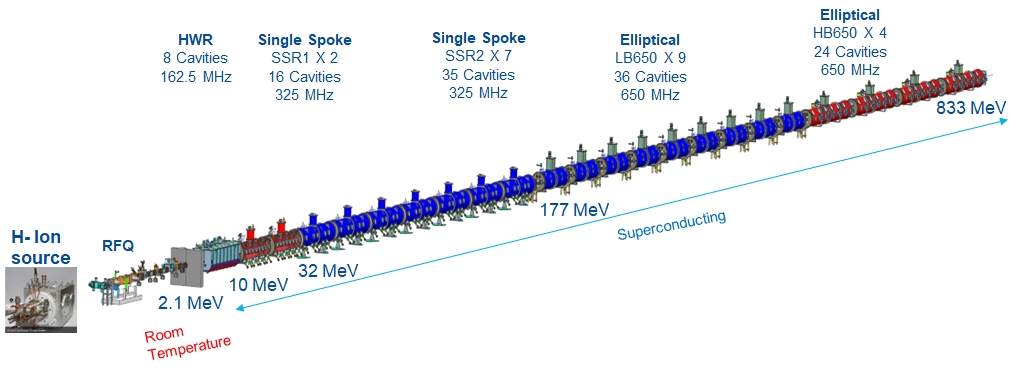}
\caption{PIP-II Linac Components}
\label {fig2}
\end{figure*}

\section{LLRF System}
\par
The PIP-II Linac is composed of seven different types of warm and superconducting RF cavities with additional differences in the resonance control systems used. Two types of LLRF system architectures and controller hardware are
used to control all the cavities. The warm front-end and the HWR cavities use a FNAL developed  LLRF controller based on the Intel Arria10 SoCFPGA[3]. The majority of the superconducting cavities use a LLRF controller based on the Marble
FPGA board using a Kintex 160 FPGA by Xilinx/AMD, developed by LBNL who are also providing the software/firmware development for these systems[4,5]. The Resonance controller is being provided by JLAB and is also based on the Marble FPGA board. The LLRF system for the superconducting cavities is designed largely based on the LCLS-II project and shares the bulk of the same codebase.The other collaborators are Lodz University of Technology(TUL) for the RFPI systems and Warsaw University of Technology(TUL) for the Master Oscillator and Precision Reference Line (MO/PRL).
\begin{figure}[!b]
\centering
 \includegraphics[width=3.0in]{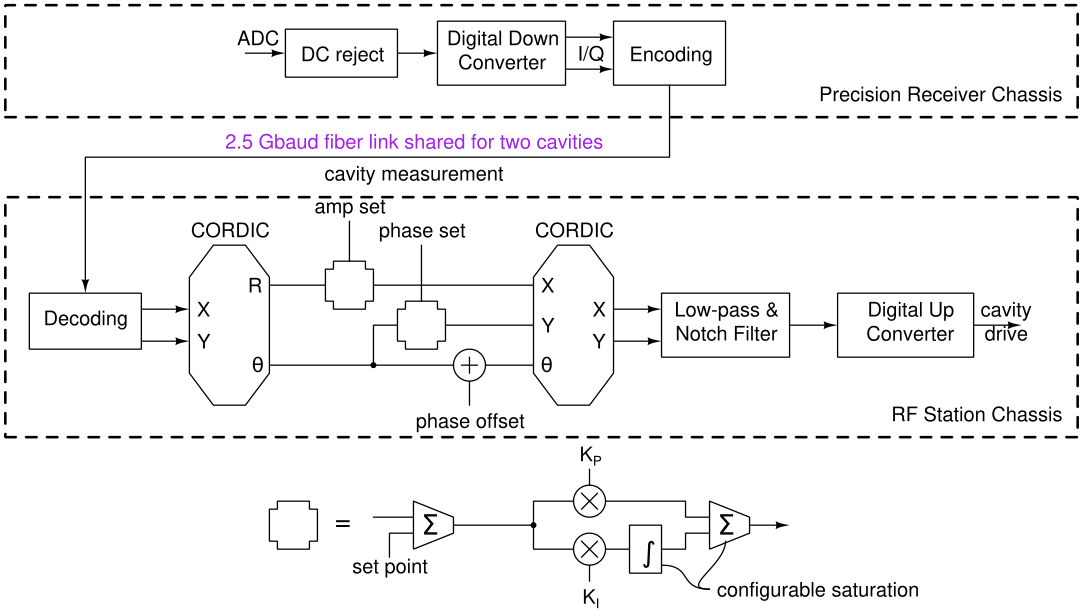}
\caption{LLRF System Architecture - Amp/Phase Control}
\label {fig3}
\end{figure}

\par
The first control system architecture, used in the warm front-end is the I/Q control loop shown in Fig.1. The I/Q control architecture has the advantage of low latency and therefore higher feedback gain for
better amplitude and phase regulation. This is the control structure implemented in the FNAL controller which also supports both CW and Pulse Mode operations. The RFQ requires pulse mode for stable field control.
\par
The second architecture used in the Marble controller is the Amplitude/Phase control loop that requires two CORDIC blocks at the input and output of the controller as shown in Fig.3. While this control scheme increases 
the latency in the feedback loop, it has several advantages that make it suitable for superconducting cavity control. The SEL architecture, with its independent amplitude and phase limiting  permits the smooth transition
between the various operating modes of operation for the cavity(SEL/SELA/SELAP) with lower transients.  It is also very effective in addressing ponderomotive oscillations as will be seen with the LB650 cavity testing discussed
later. Both architectures were tested interchangeably during the PIP2IT run and were found to perform well to meet the cavity field and phase regulation requirements. 

\section{Linac Requirements}
\par
The PIP-II project goal is to deliver 1 MW of proton beam power from the Fermilab Main Injector, over the energy
range of 60 - 120 GeV, at the start of operations of the LBNF/DUNE program. It is also a platform for eventual extension of beam power to greater than 2 MW.
The requirements for Linac beam energy stabilization are determined by the Booster RF bucket height
and requirements related to static longitudinal painting in the course of multi-turn beam injection.
The latter requires a Linac beam energy stability of 0.01 $ \% $ rms. This is achieved with the field amplitude and
phase regulation requirements 0f 0.06 $ \% $  rms and 0.06 $ ^ \circ $ rms[7]. The LLRF feedback controller
should be capable of regulating this level over a duration of less than one second. For durations
greater than one second, an adaptive feed-forward algorithm corrects for drift in the cavity phase
and amplitude calibrations by utilizing beam-based diagnostics along the Linac and in the Booster
as well as RF diagnostics to provide feed-forward corrections and slow adjustments to LLRF
calibrations.
\par
Beam current, cavity field gradients, and worst-case microphonics detuning determine loaded cavity \({Q_L}\),
bandwidths and RF power requirements. Dynamic correction of microphonic disturbances by
resonance frequency control are required for gradient regulation without exceeding available RF
power overhead. Total RF power specifications are based on cavity specifications with 2 mA of
CW beam current and 20 Hz peak amplitude microphonics. Power loss through the HWR and
SSR RF distribution is assumed to be 10\% and power loss through the 650 MHz distribution is
assumed to be 6\%. Final power specifications include a 25\% overhead in addition to distribution
loss that allows for RF control margin.
\section{Testing at CMTF and STC}
\par
PIP2IT was a test run at CMTF for the PIP-II project where the injector, warm front-end and the first two superconducting cryomodules were tested. The warm front-end consists of an Ion source, an RFQ and three buncher cavities.The superconducting RF section consists of the 8-cavity half-wave-resonator(HWR) cryomodule operating at 162.5 MHz which is followed by the 8-cavity single-spoke resonator(SSR1) cryomodule operating at 325 MHz. The LLRF systems for the RFQ and buncher 1 were based on a FPGA/DSP LLRF controller in a VXI mainframe and are being upgraded to an Arria10 SoCFPGA based controller for PIP-II. The LLRF controllers for buncher 2,3 and for both cryomodules were based on a CycloneV SOC FPGA based hardware platform -these will also be upgraded to the newer Arria10 controller.
\par
 The resonance control system for the HWR cryomodule uses a pneumatic tuner that changes the cavity resonant frequency by regulating the helium pressure on the cavity manifold. The SSR1 cavities use a piezo/stepper motor type control. The RFQ uses a cooling temperature regulation system for resonance control while the bunchers have static resonance control using stepper motors.  The data acquisition and control system can support both CW and Pulsed mode operation. Beam loading compensation is available which can be used for both manual/automatic control in the LLRF system. The user interfaces included EPICS, Labview and ACNET - they will be replaced by an entirely EPICS based user interface and control system for PIP-II. Testing of the RF system with 2 mA beam accelerated to 20 MeV has been completed in May 2021. Some of the results from the testing are shown here[2].
\subsection {RFQ and Buncher}
\par
Both the RFQ and buncher 1 LLRF systems support 20 Hz pulsed and CW mode operation. During the PIP-II-IT run only pulsed mode operation was used due to stablity issues with the RFQ. 2mA beam at 550 us pulse width was accelerated through the RF components by the end of the testing. Beam loading compensation synchronized with a beam arrival trigger was used effectively with both the RFQ and buncher cavities. The compensation is applied as an addition to the feedforward control input and supports both manual and automatic adaptive control schemes [8]. The resonance control system for the RFQ is regulated with a temperature control feedback loop and the static tuning of the bunchers was accomplished by manual stepper motor control. The RF system is turned on in self-excited loop(SEL) mode and switched to generator driven mode(GDR) with the cavities on resonance. A labview interface provides a display of all waveforms and allows for operator control of all parameters. All parameters are also available in ACNET which allows the control of the system with automated sequencers which simplifies the operator interface and also allows for data logging and monitoring.
 
\par
Beam loading compensation was implemented and tested with the buncher cavities towards the end of the run. The test was performed with 5 mA of beam with a 550 us pulse width. Manual tuning of the beam loading compensation parameters enabled the reduction of the phase disturbance at the leading edge of the beam from 3.5 degrees to \(<\) 0.2 degrees.
The relatively low Q values for the RFQ and buncher cavities do not allow large feedback gains. Table 2 shows the feedback gain limits for stable operation for the various types of cavities in the PIP-II-IT accelerator[2]. The feedback gains used were \(\approx\) 4 and 3 for the RFQ and the bunchers respectively. Loop gain measurements were performed to confirm the actual gains in the feedback operation. Testing at PIP-II-IT showed that feedback is not sufficient to compensate for beam loading and an effective BLC scheme is essential for all components in the warm front-end.
The feedback regulation performance for the warm frontend is indicated in Table 2.
\begin{table}[!t]
\caption{Cavity Maximum Feedback Gain Range}
\label{tab:sample}
\vspace{9pt}
\centering
\begin{tabular}{c c c c c}
Cavity & \( {Q_L}\) & \( f_0\) & \( f_{H} \) & \(K_P\)\\
Type& & (MHz) & (Hz) &\\
\hline \vspace{2pt}
Warm  & \(3000\) & 53 & \(8.83 \times 10^3\) & 15\\
RFQ & \(15000\) & 162.5 & \(5.542 \times 10^3\) & 23\\
Buncher & \(10000\) & 162.5 & \(8.125 \times 10^3\) & 16\\
HWR & \(2.32 \times 10^6\) & 162.5 & 35 & 3548\\
SSR1 & \(3.02 \times 10^6\) & 325 & 53.8 & 2317\\
SSR2 & \(5.05 \times 10^6\) & 325 & 32.2 & 3846\\
LB650 & \(10.36 \times 10^6\) & 650 & 31.4 & 3935\\
HB650 & \(9.92 \times 10^6\) & 650 & 32.76 & 3801\\
LCLSII & \(4 \times 10^7\) & 1300 & 16.25 & 7600\\
\end{tabular}
\end{table}

\begin{table}[!h] 
\caption{Warm Frontend Amplitude and Phase Regulation}
\label{tab:sample}
%\vspace{8pt}
\centering

\begin{center}
    \begin{tabular}{ | c | c | c | }
    \hline
\bfseries Cavity & \bfseries Amplitude ({\%} \bfseries rms) & \bfseries Phase ( \textdegree \bfseries {rms}) \\ \hline
%Cavity &  Amplitude ({\%} (rms) & Phase ( \textdegree {rms}) \\ \hline
 RFQ & 0.0863 & 0.1027  \\ \hline 
Buncher 1 &  0.0417 & 0.0356  \\ \hline
Buncher 2 &  0.0322 & 0.0209  \\  \hline
Buncher 3 &  0.0348 & 0.0253   \\ \hline
    \end{tabular}
\end{center}
\end{table}

\subsection {HWR Cryomodule}
\par
The HWR cryomodule is the first one in the superconducting section of the PIP-II LINAC. The cryomodule is comprised of 
8 cavities operating at a frequency of 162.5 MHz and accelerating beam upto 10 MeV. Resonance control of the cavities is performed with a pneumatically operated slow tuner which
compresses the cavity at the beam ports. Helium gas pressure in a bellows mounted to an end wall of the cavity is controlled by two solenoid valves, one on the pressure side and one on the vacuum side. Each tuner has
two actuators for the Pressure and Vacuum valves which are driven by two DAC channels through a signal conditioning unit. Each tuner also has a pressure transducer signal that is
digitized using a DC coupled ADC channel whose output represents the pressure in the cavity manifold which is the primary physical parameter that is controlled[10].
\par
 The resonant frequency of the cavity can be controlled in one of two modes. A pressure feedback control loop can hold the cavity tuner pressure at a fixed value for the desired resonant frequency - this is referred to as the 'Pressure Mode' operation. Alternately, the feedback loop can regulate the cavity tuner pressure to bring the RF detuning error to zero which is referred to as the 'RF Mode'. In the latter case, the tuner can compensate for slow drifts in resonant frequency. In the GDR (generator driven) mode of operation,  all cavities are run at the same reference frequency.  The RF Mode is neccessary for the GDR mode. A state machine based system controller can provide an automatic transition between the modes of operation according to the machine state. The tuner implementation is done with a signal conditioning module and the same LLRF SoCFPGA controller is used for cavity field control.Table 3 shows the field regulation performance of the HWR cavities.
\begin{table}[!h]
\caption{HWR Amplitude and Phase Regulation}
\label{tab:sample}
%\vspace{8pt}
\centering
\begin{center}
   \begin{tabular}{ | c | c | c | }
    \hline
% Cavity &  Amplitude ({\%}  rms) &  Phase ( \textdegree  {rms}) \\ \hline
\bfseries Cavity & \bfseries Amplitude ({\%} \bfseries rms) & \bfseries Phase ( \textdegree \bfseries {rms}) \\ \hline
 4 & 0.0115 & 0.0228  \\ \hline 
5 &  0.0106 & 0.0065  \\ \hline
6 &  0.0101 & 0.0056  \\  \hline
7 &  0.0081 & 0.0055   \\ \hline
8 &  0.0103 & 0.0062  \\  \hline
    \end{tabular}
\end{center}
\end{table}

\subsection {SSR1 Cryomodule}
\par
The SSR1 cryomodule consists of 8 single spoke resonator type superconducting cavities operating at 325 MHz. They are equipped with piezo tuners and stepper motors for dynamic and static resonance control. Each cavity is driven by a 7kW solid state amplifier(SSA) with a maximum operating field of 10 MV/m. The LLRF system for all eight cavities is implemented with 4 SoCMFC controller chassis and 2 resonance control chassis(RCC). The RCC is the same hardware used in the LCLS-II project with controls for the stepper motors and piezo tuners for 4 cavities per chassis. The RCC uses an EPICS user interface and has a Xilinx FPGA for the digital control and communication. The LLRF controllers are implemented with Intel SoCFPGA's with a Labview and ACNET user interface for operator control. Cavity detuning data is transmitted from each LLRF controller to its corresponding RCC over an optical fiber and high speed transcievers on both ends. This cryomodule was unique in its mixed hardware/software and user interface diversity but was integrated and made fully operational without too much difficulty.
\par
Table 4 shows the measured amplitude and phase regulation of the SSR1 cavities. All cavities meet the specifications for PIP-II.
These measurements were all taken without beam. During the short period when the full 550 us width pulse of 2 mA beam was 
tested, no beam loading effects were seen due to the high feedback gains adequately suppressing the disturbances due to beam.The microphonics detuning histograms for the HWR and SSR cavities are shown in Fig. 4. 

\begin{table}[!h]
\caption{SSR1 Amplitude and Phase Regulation}
\label{tab:sample}
%\vspace{8pt}
\centering
\begin{center}
 %   \begin{tabular}{ | p{2.0 cm}| l | l | }
   \begin{tabular}{ | c | c | c | }
    \hline
\bfseries Cavity & \bfseries Amplitude ({\%} \bfseries rms) & \bfseries Phase ( \textdegree \bfseries {rms}) \\ \hline
1 & 0.0194 & 0.0116  \\ \hline 
2 &  0.0289 & 0.0164  \\ \hline
3 &  0.0219 & 0.0118  \\  \hline
4 & 0.0157 & 0.0091  \\ \hline 
5 &  0.0140 & 0.0088  \\ \hline
6 &  0.0158 & 0.0093  \\  \hline
7 &  0.0147 & 0.0092   \\ \hline
8 &  0.0124 & 0.0076  \\  \hline
    \end{tabular}
\end{center}
\end{table}

\begin{figure}
    \centering
    \begin{subfigure}[b]{0.48\linewidth}        %% or \columnwidth
        \centering
        \includegraphics[width=\linewidth]{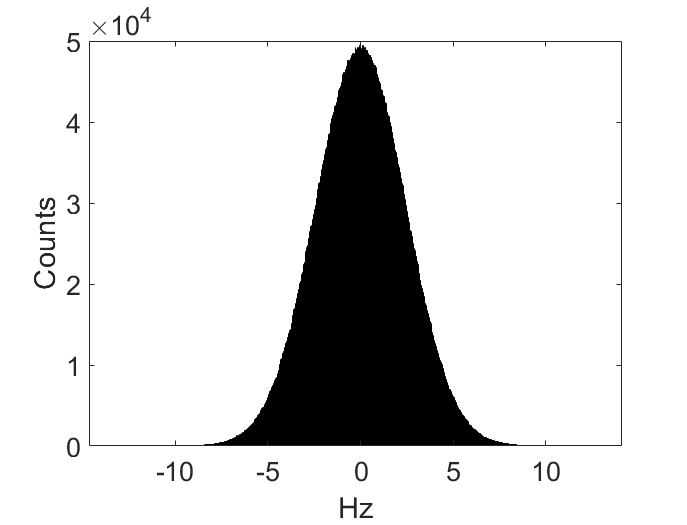}
        \caption{HWR Cavity 6}
        \label{fig:A}
    \end{subfigure}
    \begin{subfigure}[b]{0.51\linewidth}        %% or \columnwidth
        \centering
        \includegraphics[width=\linewidth]{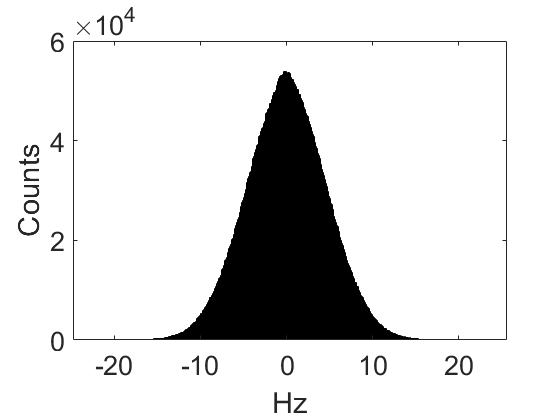}
        \caption{SSR1 Cavity 6}
        \label{fig:B}
    \end{subfigure}
    \caption{Cavity Detuning Histograms}
    \label{fig:4}
\end{figure}

\begin{figure}[!b]
    \centering
    \begin{subfigure}[b]{0.49\linewidth}        %% or \columnwidth
        \centering
        \includegraphics[width=\linewidth]{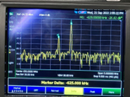}
        \caption{{\(4/5 \pi \)} Mode 625 kHz}
        \label{fig:A}
    \end{subfigure}
    \begin{subfigure}[b]{0.47\linewidth}        %% or \columnwidth
        \centering
        \includegraphics[width=\linewidth]{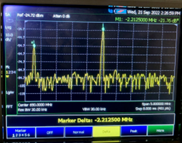}
        \caption{{\(3/5 \pi \)} Mode 2.125 MHz}
        \label{fig:B}
    \end{subfigure}
    \caption{LB650 Cavity Resonant Modes}
    \label{fig5}
\end{figure}

\subsection {HB650 and LB650 Cavity Testing at STC}
\par
Single cavity testing of various PIP-II cavities has been an ongoing effort at STC facility since 2019. HB650, LB650 and SSR2 cavities have been tested here. It has proved to be
a good test bed for testing and validating various LLRF system hardware components for PIP-II. The Jlab provided Resonance Controller, LBNL's LLRF controller along with the FNAL
designs of LLRF controllers have all been tested here.
\par
 The first LB650 cavity was tested last year with significant ponderomotive force effects that highlighted the advantages of the SEL
architecture. The five cell cavity had significant excitation of the  \({4/5} \pi \) and  \({3/5} \pi \) passband modes as shown in Fig. 5. These modes had to be suppressed with notch filters before increasing the
feedback gains and reaching the full gradient of 17 MV/m in GDR mode. The cavity waveforms are shown in Fig. 6.

\begin{figure}[!t]
\centering
 \includegraphics[width=3.0in]{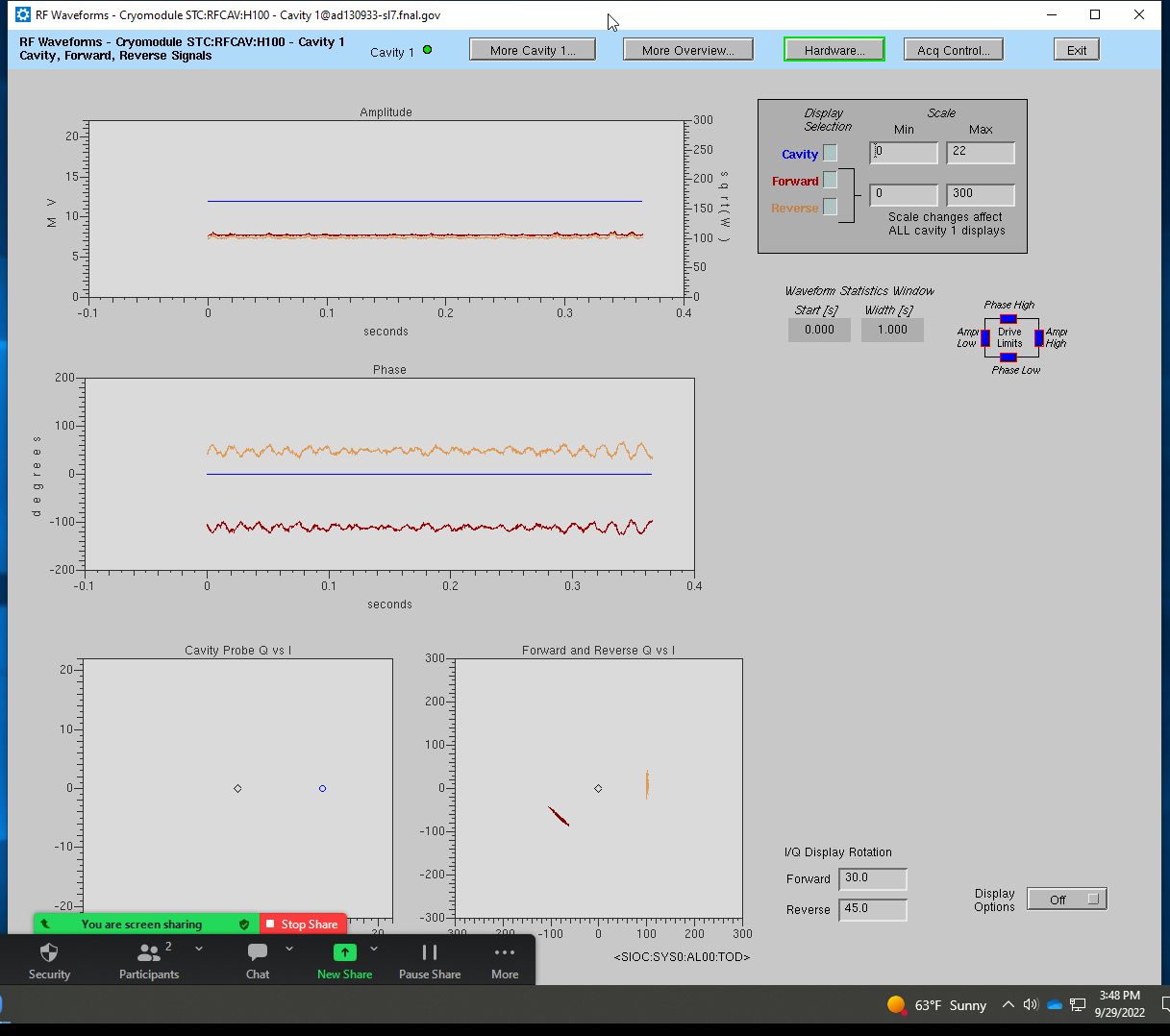}
\caption{LB650 Cavity Waveforms in GDR Mode}
\label {fig6}
\end{figure}

\begin{figure}[!b]
\centering
 \includegraphics[width=3.0in]{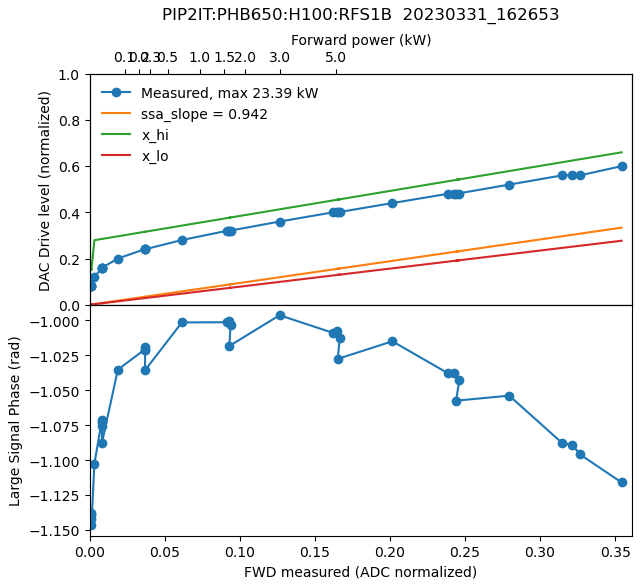}
\caption{HB650 SSA Calibration- Affine Fit }
\label {fig7}
\end{figure}

\subsection {HB650 Test Stand at CMTF}
\par
A prototype HB650 cryomodule with 6 cavities has been under test at the CMTF facility since January 2023. A single 40 kW SSA provided by DAE, India has been used
to test the cavities, one at a time. This facility will be used to test all the PIP-II cryomodules that have not been tested - HB650, LB650 and SSR2. This facility is also an 
important test bed to validate the performance of various subsystems for PIP-II such as SSA's, RFPI, MPS, Timing and Synchronization and the EPICS controls interface. Additional testing with the prototype HB650 cryomdule will resume when it returns from its transportation test to the UK at the end of this year. The amplifier non-linearities have been a challenge to run the cavity in GDR mode. SSA calibration procedures and computations were modified to compensate for
the non-linearities of the SSA as shown in Fig. 7.
\par
The PIP-II RFPI prototype hardware designed by TUL was successfully tested and has entered the final design prototype phase.
New LLRF and Resonance control hardware based on the Marble FPGA board are being readied for testing at CMTF. 

\section{Conclusion}
\par
The PIP-II LLRF system  components are in the final design phase. Testing at the two teststands STC and CMTF over the past few years has provided the opportunity to test and validate both hardware and
software/firmware components of the design with all cavity types with and without beam. These facilities will continue to serve as test areas where additional cryomodules and cavities will be tested
until installation and commissioning starts at the end of 2026. The control system and  user interfaces implemented with EPICS will also get tested in these test stands.
\par
Resonance control with temperature control, helium pressure control and piezo tuners were tested in the PIP2IT phase. The project specifications for microphonics detuning were met by all
the cavities tested so far. Field magnitude and phase regulation specifications for these LLRF systems for the project have also been met. The phase reference line components and timing
and synchronization subsystems will also be tested here as they become finalized.

\end{document}